\documentclass{ifacconf}

\usepackage{graphicx}      
\usepackage{natbib}

\usepackage{lipsum}
\usepackage{amsmath,amssymb,amsfonts,bm}
\usepackage{breqn}
\usepackage[T1]{fontenc}
\usepackage{xcolor}
\usepackage{appendix}
\usepackage{booktabs}
\usepackage{algorithm}
\usepackage{algpseudocode}
\usepackage{adjustbox}
\usepackage[normalem]{ulem}
\usepackage{xparse}
\usepackage{ragged2e}
\usepackage{changepage}

\usepackage{tikz}
\usetikzlibrary{shapes,arrows,shadows,positioning,calc,math}
\usetikzlibrary{decorations.text}
\usetikzlibrary{backgrounds}
\usepackage[siunitx]{circuitikz}

\newcommand{\norm}[1]{\left\lVert#1\right\rVert}
\begin{document}
\begin{frontmatter}

\title{Real-time Battery State of Charge and parameters estimation through Multi-Rate Moving Horizon Estimator} 


\author[1]{Tushar Desai} 
\author[2]{Federico Oliva}
\author[1]{Riccardo M.G. Ferrari}
\author[2]{Daniele Carnevale}

\address[1]{Delft Center for Systems and Control, Delft University of Technology, Delft, Netherlands (e-mail: t.k.desai@tudelft.nl, r.ferrari@tudelft.nl).}
\address[2]{Department of Civil Engineering and Computer Science Engineering, University of Rome "Tor Vergata", Rome, Italy. Email: federico.oliva@students.uniroma2.eu,daniele.carnevale@uniroma2.it}
\thanks[]{This work has been partially supported by the Italian Ministry for Research in the framework of the 2020 Program for Research Projects of National Interest (PRIN). Grant No. 2020RTWES4.}


\begin{abstract}                
For reliable and safe battery operations, accurate and robust State of Charge (SOC) and model parameters estimation is vital. However, the nonlinear dependency of the model parameters on battery states makes the problem challenging. We propose a Moving-Horizon Estimation (MHE)-based robust approach for joint state and parameters estimation. Dut to all the time scales involved in the model dynamics, a \textit{multi-rate} MHE is designed to improve the estimation performance. Moreover, a parallelized structure for the observer is exploited to reduce the computational burden, combining both \textit{multi-rate} and a reduced-order MHEs. Results show that the battery SOC and parameters can be effectively estimated. The proposed MHE observers are verified on a Simulink-based battery equivalent circuit model.
\end{abstract}


\end{frontmatter}
\section{Introduction}
\label{scn:Introduction}
Lithium-ion batteries have become popular in various applications due to their higher energy density and longer cycle life. Accurately determining the \textit{State of Charge} (SOC) is crucial for safe and reliable battery operations. However, SOC cannot be measured directly but can be estimated through nonlinear transformations using battery monitoring signals and models.\\
Model-based state estimation techniques use three main types of modeling methods: Electro-Chemical models (EC), Equivalent Circuit Models (ECM), and Data-driven models. EC models are highly accurate but computationally expensive, making them challenging for onboard battery management. On the other hand, modern data-driven models based on machine-learning can capture adequately nonlinearities but may not generalize well beyond the training dataset.\\
ECM, instead, consists of a lower-order battery model with linear state dynamics, striking a good balance between accuracy and computational complexity. 
ECMs have an output nonlinearity due to the Open circuit Voltage (OCV)-SOC nonlinear mapping and model parameters have a nonlinear and time-varying dependency on SOC, temperature, and battery health \citep{Tran2021}. Thus nonlinear extensions of Kalman filters, such as Extended \citep{Plett2004}, Sigma-point \citep{Plett2006} and Cubature \citep{Peng2019} are widely used as a joint/dual approach to estimate SOC and parameters. These filters provide good estimation accuracy and efficiency.\\
However, Kalman family algorithms have some fundamental limitations \citep{Shrivastava2019}: Their accuracy depends on the initialization of the SOC, parameters, and covariance matrix elements. Also, the Kalman filters cannot handle constraints on the parameters and states. Other nonlinear observers, such as Moving Horizon Estimators (MHE) and High-gain observers \citep{Carnevale}, have been exploited to cope with these limitations. MHE methods optimize over a receding horizon data window rather than a single instantaneous measurement. This approach provides more robust and smooth estimates under modeling uncertainties and disturbances \citep{Haseltine2004}. \\
\cite{Hu2018} implemented an MHE with an EC model variant, the Single particle model (SPM). \cite{Yan2021} presented an MHE implemented on a first-order ECM model, whose parameters' dependency on the SOC was evaluated offline. \cite{shen2016online,shen2018accurate} used an MHE based adaptive observer to estimate SOC and battery parameters.\\
Indeed, this work builds on the results presented by Shen et al., which compares a joint-MHE algorithm with the standard EKF approach. The general idea is to improve the MHE performance and robustness by addressing the problem of the best choice of the observer horizon, exploiting the signal variability analysis proposed in the \textit{adaptive} MHE by the authors \citep{Oliva01,Oliva02}. The online parameter estimation is considered to track the slow time-varying dynamic of the battery. The first contribution of this paper is a \textit{multi-rate} version of the MHE. We show that this approach improves the estimation of plant parameters across different time scale dynamics, at the cost of a significant computational burden in the algorithm. Indeed, the second goal of this work is to move towards a real-time implementation of the \textit{multi-rate} MHE. Thus, a different approach is presented, namely the \textit{parallel} MHE. This method exploits two concurrent MHEs to provide a local SOC estimation and a more general model for the battery parameters. This last approach could come in handy if both model characterization and fault detection were to be achieved on a plant. All the methods above are tested on a Simulink-based test bench.\\
The rest of the paper is divided into six sections as follows. The primary SOC-dependent ECM model is detailed in Sec. \ref{scn:model}. The MHE framework is defined in Sec. \ref{scn:MHE_intro}, and the signal variability and buffer length choices are introduced in Sec. \ref{scn:rich_adapt}. Sec. \ref{scn:multiscaleMHE} entails the \textit{multi-rate} MHE. Lastly, the \textit{parallel} MHE solution is described in Sec. \ref{scn:parallelMHE}. Conclusion and future works are presented in Sec. \ref{scn:conclusion}.
\section{ECM model with SOC dependency}
\label{scn:model}
\begin{figure}[ht!]
\begin{center}
    \begin{circuitikz}[scale = 0.8, transform shape]
        \draw (0,0) to[battery,l=$V_{\text{OCV}}(Z)$] (0,3) to[R, l=$R_0(Z)$,i_=I] (3,3);
        \draw (3,3) -- (3.5,3);
        \draw (3.5,2) -- (3.5,4);
        \draw (3.5,4) to[R=$R_1(Z)$] (5.5,4); 
        \draw (3.5,2) to[C=$C_1(Z)$,v_=$V_1$] (5.5,2);
        \draw (5.5,2) -- (5.5,4);
        \draw (5.5,3) -- (7,3) to[voltmeter, l=$V_b$] (7,0)  -- (0,0);
    \end{circuitikz}    
\end{center}
\caption{\label{fig:first_order_ECM}\centering First-order ECM of battery }
\end{figure}
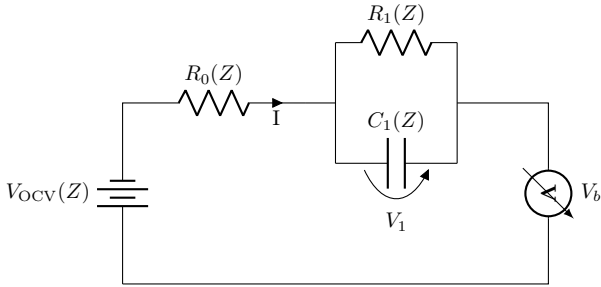

\vspace{-3pt}
Considering the trade-off between accuracy and computation time, the first-order battery ECM shown in Fig. \ref{fig:first_order_ECM} is considered here \citep{Hu2012}. $V_{\text{OCV}}(Z)$ is the Open circuit voltage dependent on the state of charge $Z$. $R_{0}$ is the ohmic resistance. $R_{1}$ and $C_{1}$ are the polarization resistance and polarization capacitance respectively. The load current, denoted by $I$, is the plant's exogenous input. $V_{b}$ is the terminal voltage when the load is applied, and is assumed the measured output. The dynamic voltage across the RC pair is denoted by $V_{1}$. 

\vspace{-6pt}

\begin{figure}[h!]
    \centering
    \includegraphics[width=8cm]{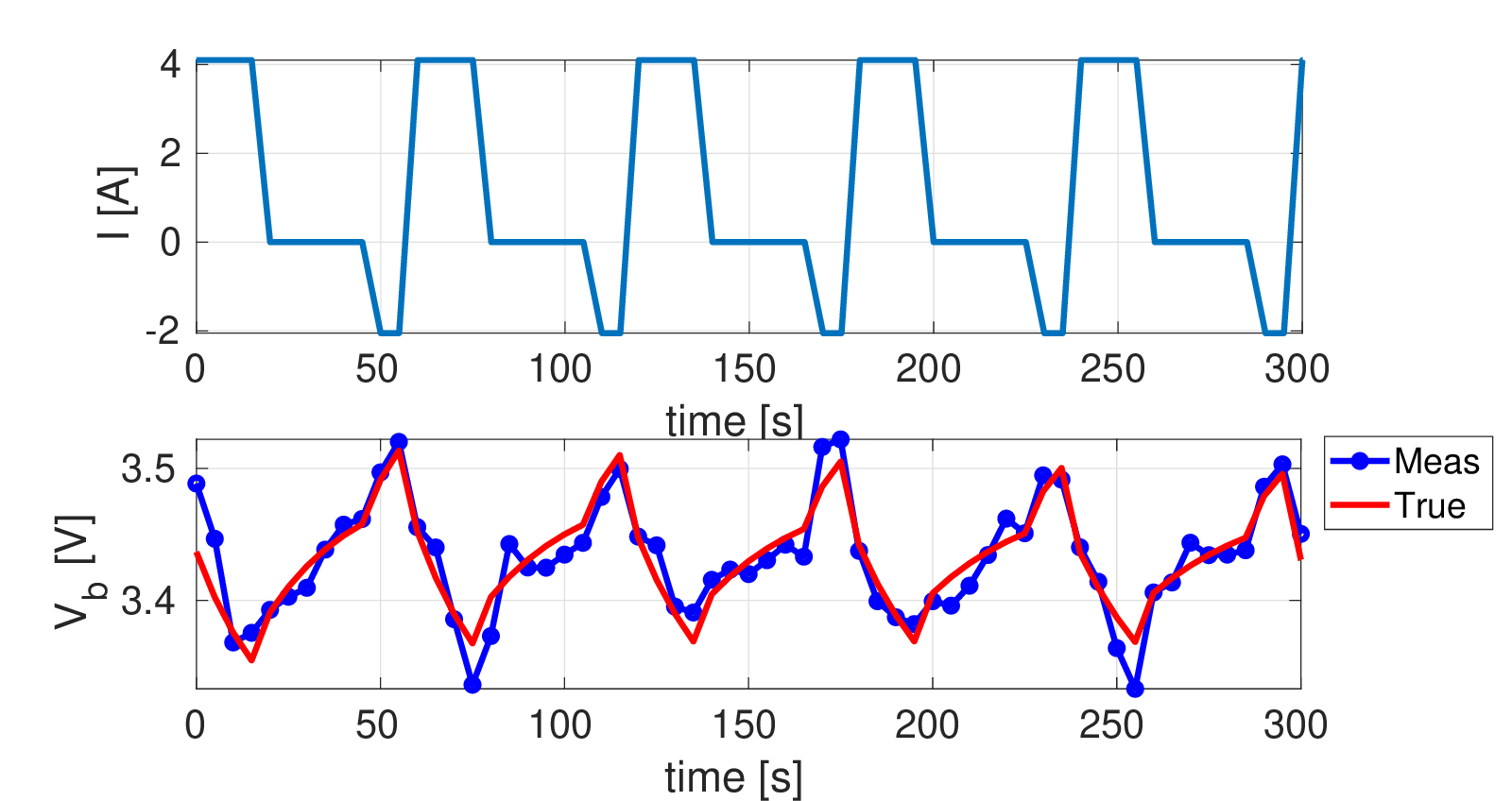}
    \caption{Output measurements $V_b$ related to the modified HPPC cycle.}
    \label{fig:ynoiseAndInput}
\end{figure}

The ECM dynamics can be written in discrete time as: 

\begin{subequations}
\label{eqn:first_order_ECM_equations}
\footnotesize
\begin{align}
    Z_{k+1} &= Z_{k} - \dfrac{\eta T_{s} I_{k}}{C_n}\\
    V_{1,k+1} &= \exp \left(-\dfrac{T_s}{R_1(Z_k) C_1(Z_k)}\right) V_{k} \notag \\
    &+\left(1 - \exp \left(-\dfrac{T_s}{R_1(Z_k) C_1(Z_k)}\right)\right) R_1(Z_k)I_{k}  \\
    V_{b, k} &= V_{\text{OCV}}(Z_k)-I_{k}R_0(Z_k)-V_{1,k}
\end{align}
\normalsize
\end{subequations}

where $C_{n}$ denotes the nominal battery capacity, and $\eta$ is the coulombic efficiency. $T_{s} = 1s$ is the system sampling time. The temperature is kept constant for this experiment at $T=313.5 \text{ [K]}$. The nonlinear dependency of $V_{\text{OCV}}, R_0, R_1, C_1$ on $Z$ is measured offline, and it is assumed to be polynomial as described below

\begin{subequations}
\label{eqn:ECM_parameters_and_SOC}
\begin{align}
   V_{\text{OCV}}(Z) &= \sum_{i=0}^{P_{ocv}}\alpha_{V_{OCV, i}} \cdot Z^{i},\\
   R_{0}(Z) &= \sum_{i=0}^{P}\alpha_{R_{0, i}} \cdot Z^{i},\\
   R_{1}(Z) &= \sum_{i=0}^{P}\alpha_{R_{1, i}} \cdot Z^{i},\\
   C_{1}(Z) &= \sum_{i=0}^{P}\alpha_{C_{1, i}} \cdot Z^{i},
\end{align}
\end{subequations}

where $\alpha_{\star}$ are the polynomial coefficients. $P$ and $P_{ocv}$ are the order of the polynomials, with $P_{ocv}$ generally higher than $P$. To ease the overall analysis, no process noise is considered in the model, as in \cite{shen2018accurate}.\\
Regarding the input current, we modified the standard Hybrid Pulse Power Characterization (HPPC) method for input current by using a 20s 1C discharge rate current, 30s rest, and 10s 0.5C charge to evaluate battery parameter dependencies on different SOC levels.  All the experiments are carried out on the Simulink-based parameterized ECM test bench to evaluate the performance of our approaches \citep{matlab_batterytablebased}. Moreover, the measurement signal has been affected by Gaussian noise distributed as $\mathcal{N}(0, 0.05)$. A number of five HPPC periods and the related true and noisy measurements are depicted in Fig. \ref{fig:ynoiseAndInput}. 

\begin{figure}[h!]
    \centering
    \includegraphics[width=8cm]{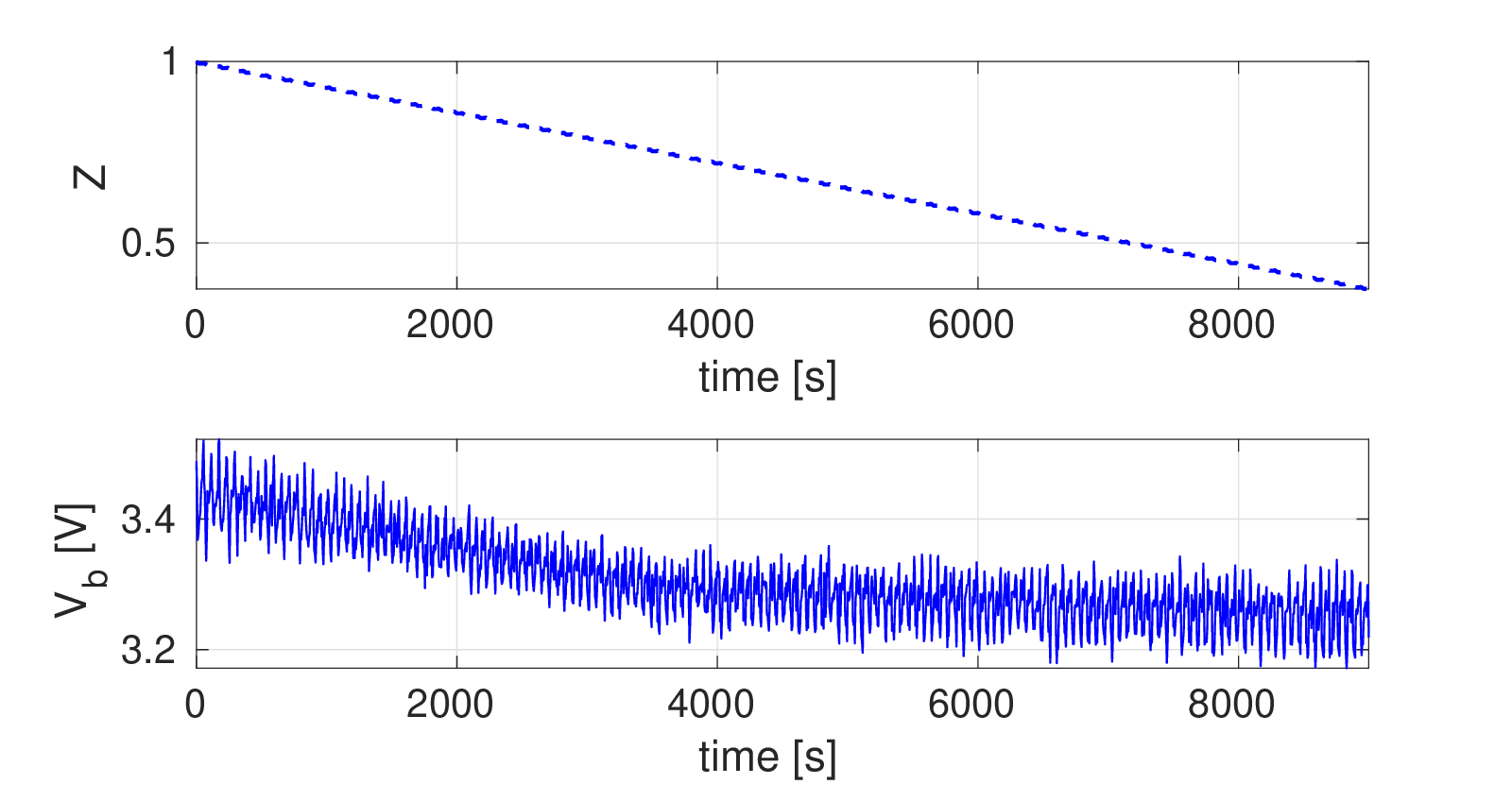}
    \caption{\centering True SOC and $V_b$ evolution.}
    \label{fig:SOC&VbEvolution}
\end{figure}

The SOC and $V_b$ time evolution considered in this paper are reported in Fig. \ref{fig:SOC&VbEvolution}.

\section{Moving-Horizon Estimator}
\label{scn:MHE_intro}
This section briefly recalls the definition of \textit{Moving-Horizon Estimators} (MHEs), along with the \textit{adaptive} versions proposed in \cite{Oliva01,Oliva02}. 

\subsection{Standard MHE}
\label{subscn:MHE_standard}
Consider a general nonlinear system in the following form:

\begin{align}
	\label{eqn:conv_general_system}
	\mathcal{P}: \
	\begin{cases}
		\dot{\bm{\xi}} & = f(\bm{\xi},\bm{u},\bm{\Theta}),  \\
		\bm{y}         & = h(\bm{\xi},\bm{u},\bm{\Theta}),
	\end{cases}
\end{align}

where $\bm{\xi}\in\mathbb{R}^n$ is the state vector, $\bm{u}\in \mathbb{R}^m$ is the control input, and $\bm{y} \in \mathbb{R}^p$ is the measured output. $\bm{\Theta}$ is a vector of model constant parameters, i.e. $\dot{\bm{\Theta}} = 0$. The system is assumed to be $(N+1)$-step observable following the definition in \cite{Wynn2014}. Consider also the $N$-lifted system $\text{H}_k(\bm{\xi,\Theta})$ at time $t_k$ obtained from eq. \eqref{eqn:conv_general_system} following the definition in \cite{Tousain}. Moreover, we define $\text{Y}_k$ as the buffer containing $N$ samples of the output $\bm{y}$ starting from $t_{k,0}=t_{k-N+1}$, namely the time instant of the first (oldest) element in the buffer. The MHE problem consists of a nonlinear observer solving the following optimization problem in the estimated variables $(\hat{\bm{\xi}}_{k,0},\hat{\bm{\Theta}}_{k,0})$

\large
\begin{dmath}
	\underset{(\hat{\bm{\xi}}_{k,0},\hat{\bm{\Theta}}_{k,0})}{\text{min}} \ J = \ \text{W}_1\underbrace{||\text{Y}_k-{\hat{\text{H}}_k}||}_{\text{\textbf{output mismatch}}} + \underbrace{\text{W}_2||\hat{\bm{\xi}}_{k,0} - \hat{\bm{\xi}}^0|| + \text{W}_3||\bm{\hat{\Theta}}_{k,0} - \hat{\bm{\Theta}}^0||}_{\text{\textbf{arrival cost}}},
	\label{eqn:minproblem_normalised}
\end{dmath}
\normalsize

where $\text{W}_i$ are scaling factors specifically designed to normalize all the terms in the cost function $J$, and $(\hat{\bm{\xi}}^0,\hat{\bm{\Theta}}^0)$ are the estimated values at the beginning of the optimization process. \\
We refer to this formulation of the MHE as single-shooting MHE (\cite{Kang2006,Kuhl2011}). As far as the model integration is concerned, we refer to the same setup considered in \cite{Oliva01}, namely $T_s$ is the sampling time, while the signals considered in the measurement buffer are down-sampled every $N_{T_s}\cdot T_s$ seconds. $N_{T_s}$ is the down-sampling factor. In this paper, to solve the optimization problem in \eqref{eqn:minproblem_normalised}, the  MATLAB built-in simplex-like \textbf{fminsearch} optimization algorithm has been considered, with a fixed number of iterations $K=1$. The number of iterations is usually limited to speed up the estimation process, even though this implies a sub-optimal observer. Results in terms of local and practical stability have been introduced in \cite{Kuhl2011} and \cite{Wynn2014} stating the necessity of the arrival cost term in the minimization. 

\subsection{Adaptive MHE}
\label{subscn:adapt_MHE}

The main issue with MHEs is the computational cost which increases with the buffer length due to the evaluation of $\text{H}_k(\hat{\bm{\xi}}_{k,0},\hat{\bm{\Theta}}_{k,0})$ in \eqref{eqn:minproblem_normalised} integrating the plant dynamics in \eqref{eqn:conv_general_system}\footnote{LTI systems integration can be significantly accelerated through closed-form solutions of the ODEs.}. Recall that $N\geq n$, where $n$ is a model order, is a necessary condition on the buffer length to obtain a local convergence of the estimation (\cite{Kang2006}), while \cite{Aeyels} proves $N\geq 2n+1$ to be a sufficient condition for the estimation error convergence. Therefore, computational cost increases with the model dimension. A possible approach to increase the efficiency of MHEs relies on the \textit{signal variability} defined in \cite{Oliva02}, namely an index describing how much a measurement is informative with respect to the ones already stored in the buffer $\text{Y}_k$. The down-sampling factor $N_{T_s}$ can be changed at run-time to improve the MHE computational cost and the estimation performance by maximizing such an index. The usage of the \textit{signal variability} index will be exploited in this work to ensure the estimation's robustness and make a feasible real-time implementation of the observer. 
\section{Signal variability on Standard MHE}
\label{scn:rich_adapt}

This section investigates the role of signal variability in the MHE implementation and, more specifically, in the measurement buffer selection.

\subsection{Signal variability}
\label{subscn:signal_richness}
At each time $t_k$ of the trajectory, we define the signal variability $\delta_y$ as follows

\begin{figure}[h!]
    \centering
    \includegraphics[width=8cm]{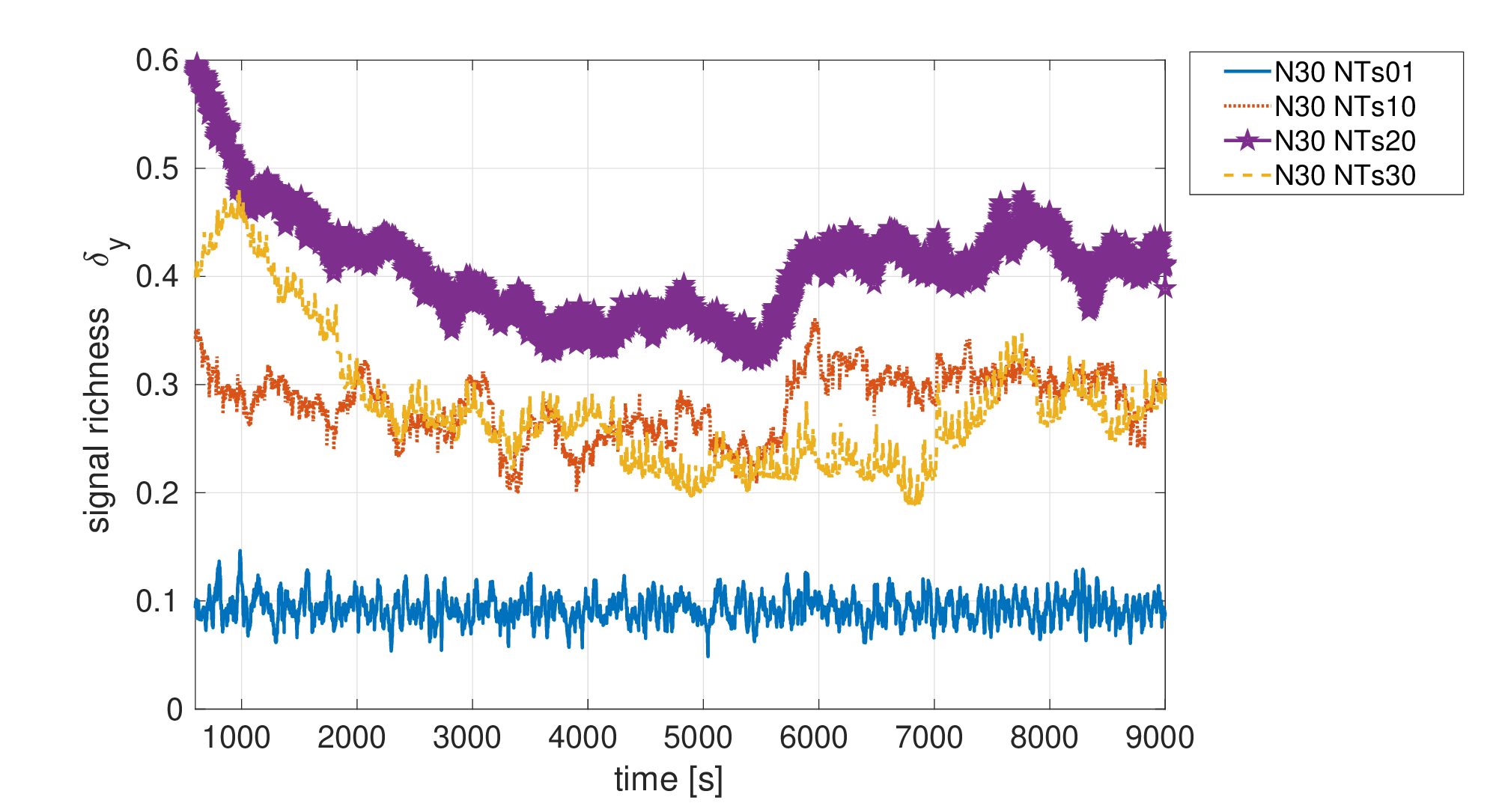}
    \caption{\centering Comparison of $\delta_y$ depending on $N_{T_s}$.}
    \label{fig:SignaRichnessComparison}
\end{figure}

\vspace{-2pt}
\begin{align}
      \label{eqn:delta_y}
      \delta_y(k,q,N_{T_s}) = & \sum\limits_{i=1}^{N-1}
      \norm{y_{k-iN_{T_s}} - y_{k-(i+1)N_{T_s}}} + \nonumber \\ 
      &\norm{y_{k+qT_s}-y_{k-N_{T_s}}} ,
\end{align}

with $q\in[0,N_{T_s}-1]$. Note that $\delta_y$ considers the buffer elements. Still, it is evaluated at each $y_k$, namely $\delta_y$ is a measure of the signal variability over the time window, and it is affected by the choice of $N_{T_s}$. Fig. \ref{fig:SignaRichnessComparison} shows $\delta_y$ over the whole output trajectory in Fig. \ref{fig:SOC&VbEvolution}, and considering different down-sampling values $N_{T_s}$, while keeping $N=30$\footnote{The period considered starts after all the measurement buffers have been filled with data.}. 
Indeed, as discussed in \cite{Oliva01}, the selection of the down-sampling factor $N_{T_s}$ improving the estimation performance coincides with the one maximizing $\delta_y$. Thus, by looking at the results in Fig. \ref{fig:SignaRichnessComparison}, the down-sampling factor is set to $N_{T_s} = 20$.

\subsection{Numerical results}
\label{subscn:adaptive_results}

\begin{figure}[h!]
    \centering
    \includegraphics[width=9.5cm]{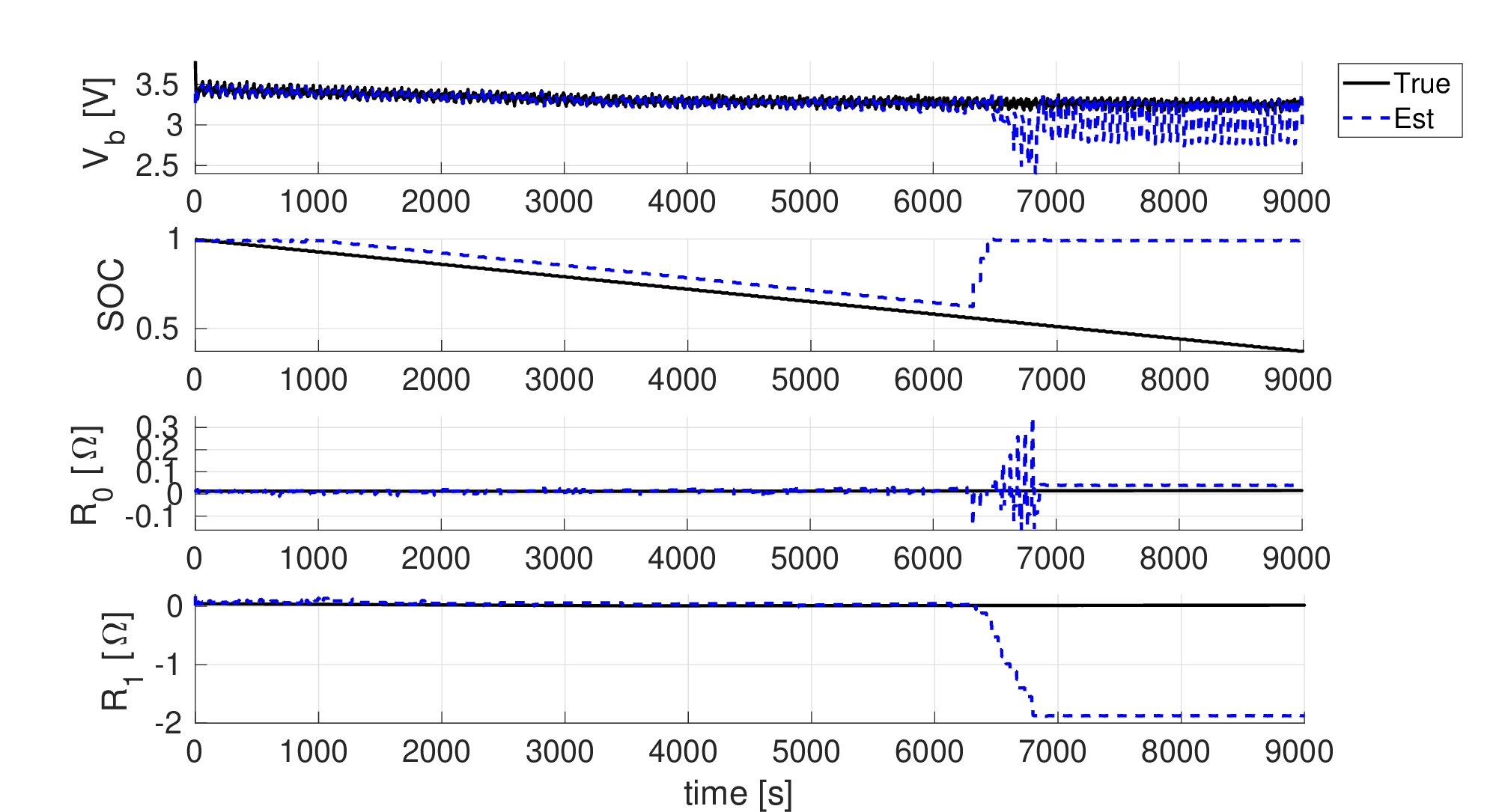}
    \caption{$V_b, Z, R_0(Z), R_1(Z)$ estimation with $N=30,N_{T_s}=1$.}
    \label{fig:R1_interp}
\end{figure}

This section compares the estimation performance depending on the $N_{T_s}$ down-sampling selection. Indeed, the variation of the measured output $V_b$ related to the SOC has relevant changes in a period of around 1000s, i.e., we can say that it changes slowly with respect to the sampling time $T_s$. Thus, by considering a low $N_{T_s}$, the measurement buffer $\text{Y}_k$ is related to very similar values of $Z$, i.e., it contains low variability of the signal $Z$, which deteriorates the estimation capability. In particular, the higher-order terms in the polynomial coefficients $(\alpha_{R_0,0-3},\alpha_{R_1,0-3},\alpha_{C_1,0-3})$ would be tough to estimate since the sensitivity of $\text{Y}_k$ to them is negligible in such a small range of $Z$ changes, as also discussed in \cite{shen2018accurate}. \\
We show now that this issue is related to the down-sampling choice, as discussed in Sec. \ref{subscn:signal_richness}. Consider the \textit{standard} MHE in eq. \eqref{eqn:minproblem_normalised} where the estimated variables are defined as $\bm{\xi} = (Z,V_1)$ and $\bm{\Theta} = (\alpha_{R_0,0-3},\alpha_{R_1,0-3},\alpha_{C_1,0-3})$. The buffer setup considers $(N=30,N_{T_s}=1)$, where $N \geq 2(n+\text{dim}(\bm{\Theta}))+1 = 29$  accordingly to \cite{Aeyels}. The $N_{T_s} = 1$ selection results in poor estimation performance, as reported in Fig. \ref{fig:R1_interp}, where the blue dashed lines show the estimated values for $(V_b, Z, R_0(Z), R_1(Z))$, compared with the true values in solid black. 

As noted by \cite{shen2018accurate}, the estimation of $V_b$ suffers from numerical instabilities: by inspecting \eqref{eqn:first_order_ECM_equations}, it is evident that such instabilities occur as  $(R_0(Z), R_1(Z))$ approach zero (numerical instability) or even negative (model \eqref{eqn:first_order_ECM_equations} instability). The effects of such instabilities are shown in \ref{fig:R1_interp}, right after $t=6000$ s, where the state estimation diverges as soon as the optimization process returns negative values for $\text{R}_1$. Such a situation is physically infeasible. A first solution to this issue consists in adding a barrier term in the cost \eqref{eqn:minproblem_normalised}, penalizing $\bm{\Theta}$ values yielding $(R_0(Z), R_1(Z), C_1(Z))$ close to 0. This first solution has been implemented in this work modifying $J$ as 

\large
\begin{dmath}
	\underset{(\hat{\bm{\xi}}_{k,0},\hat{\bm{\Theta}}_{k,0})}{\text{min}} \ J = \ \text{W}_1||\text{Y}_k-{\hat{\text{H}}_k}|| + \\ \text{W}_2||\hat{\bm{\xi}}_{k,0} - \hat{\bm{\xi}}^0|| + \text{W}_3||\bm{\hat{\Theta}}_{k,0} - \hat{\bm{\Theta}}^0|| + \\ \\ \text{b}_{R_0}(\hat{\bm{\Theta}}_{k,0}) + \text{b}_{R_1}(\hat{\bm{\Theta}}_{k,0})
	\label{eqn:minproblem_barrier}
\end{dmath}
\normalsize
where $\text{b}_R$ is a barrier function defined as follows
\begin{equation}
    \text{b}_R = 
    \begin{cases}
    0 & \text{if } R\geq 0, \\
    M & \text{otherwise},
    \end{cases}
\end{equation}

\medskip
with $M\geq\Bar{M}$ sufficiently large. Indeed, the choice of $\Bar{M}$ depends on the order of magnitude of the cost function. A rule-of-thumb consists in choosing $\Bar{M}$ several times the order of magnitude of $J$. In our case, we chose $\Bar{M}=10E5$.  Note that in this work, any algorithm solving \eqref{eqn:minproblem_barrier} with fixed $N_{T_s}$ will be referred to as \textit{standard MHE}. Indeed, the related sampling $N_{T_s}$ will always be explicit to avoid ambiguities.

\begin{figure}[h!]
    \centering
    \includegraphics[width=8cm]{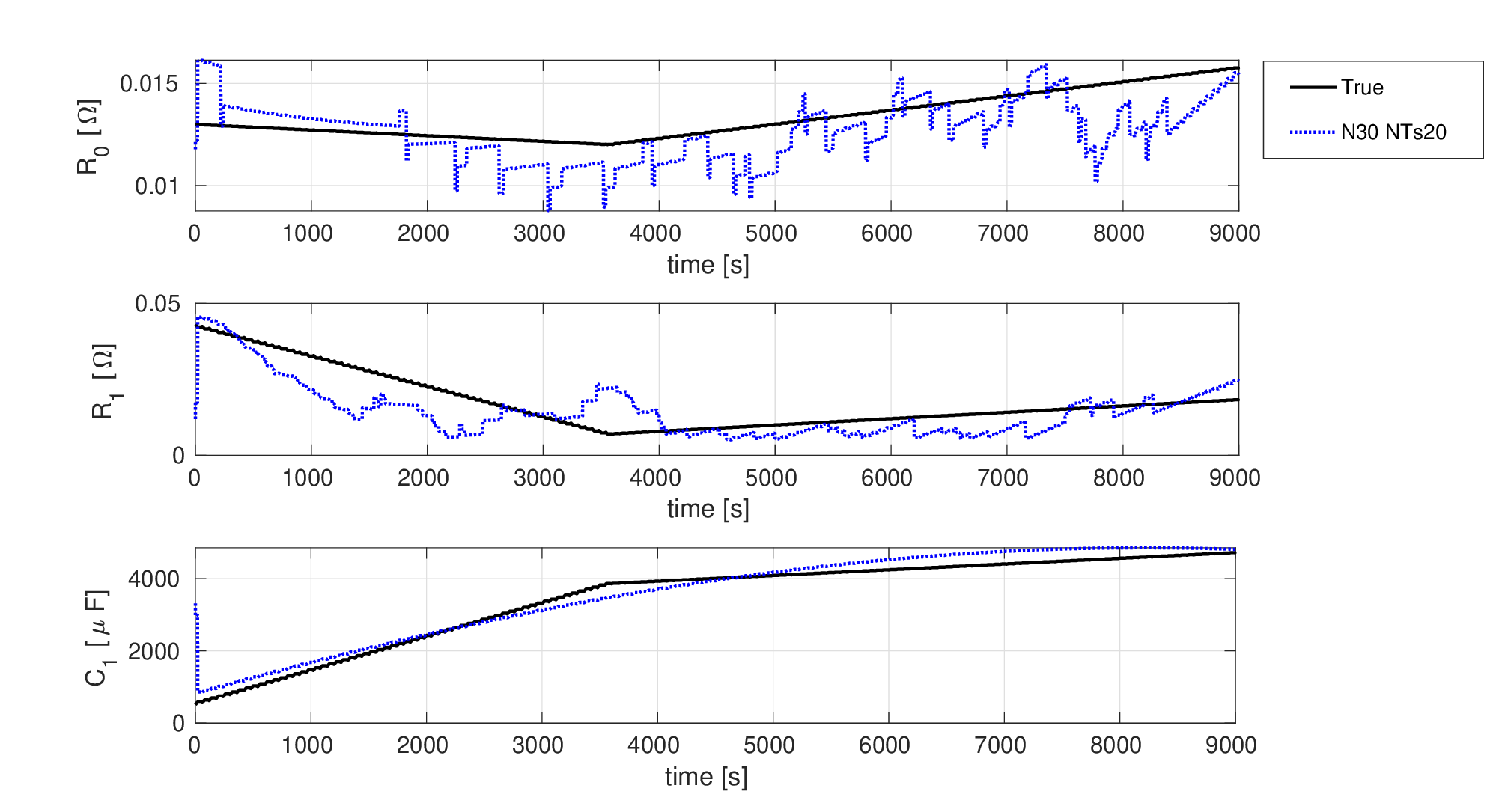}
    \caption{\centering Parameter estimation with $N=30,N_{T_s}=20$.}
    \label{fig:paramsCompare}
\end{figure}

\begin{figure}[h!]
    \centering
    \includegraphics[width=8cm]{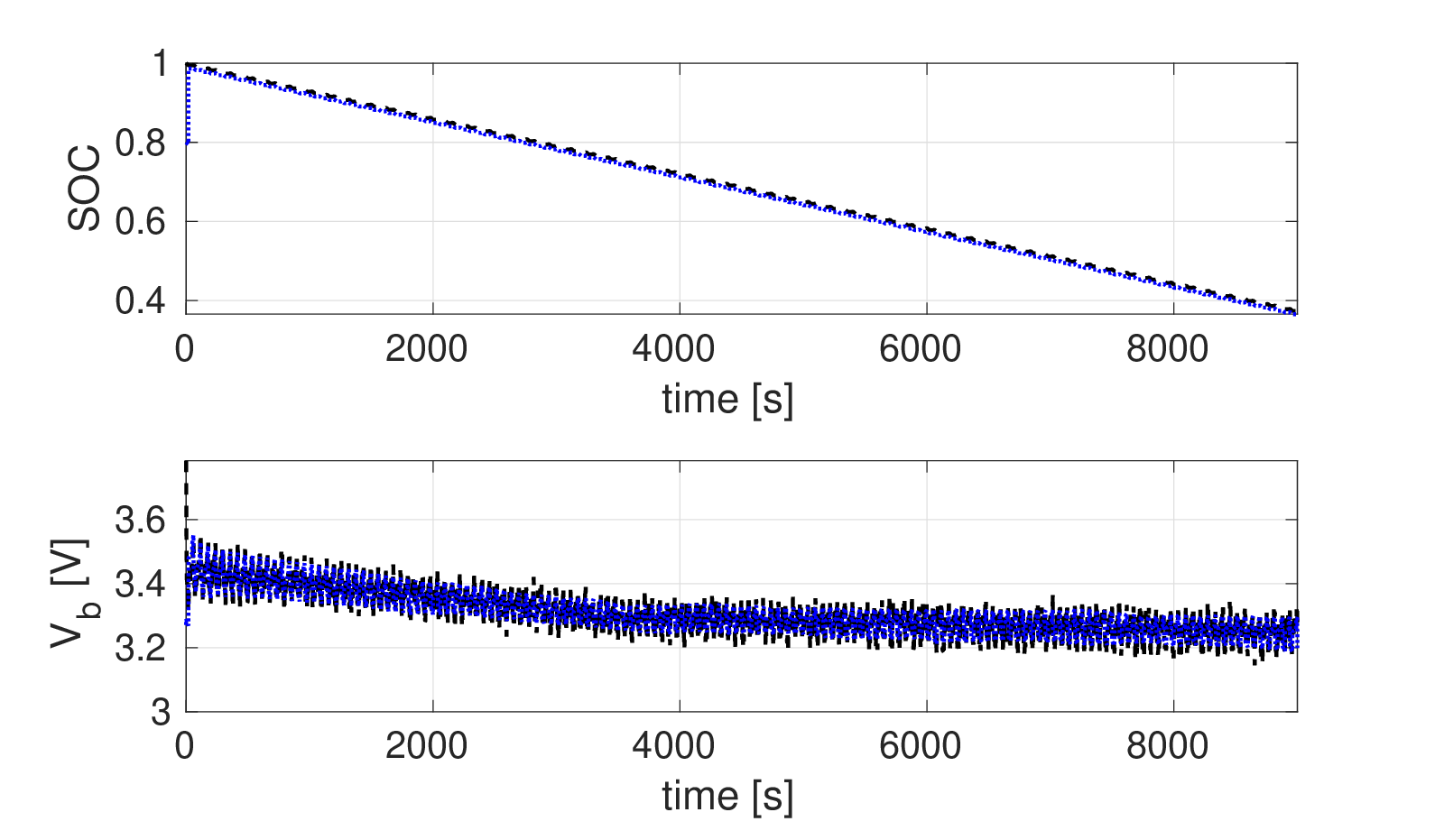}
    \caption{\centering SOC and $V_b$ estimation with $N=30,N_{T_s}=20$.}
    \label{fig:3020_SOCVb}
\end{figure}

\smallskip
When  the optimization variables are restricted to $\bm{\xi} = (Z, V_1)$ and $\bm{\Theta} = (\alpha_{R_0,0},\alpha_{R_1,0},\alpha_{C_1,0})$, i.e., the higher-order terms in the polynomial are kept constant (initial guess), the SOC and parameters estimation presents a drift as can also be seen in \cite{shen2018accurate}. Fig. \ref{fig:paramsCompare} and Fig. \ref{fig:3020_SOCVb} show the parameters, SOC, and $V_b$ estimation results for the $(N=30,N_{T_s}=20)$ case. Note that these results have been obtained by solving a \textit{standard} MHE on \eqref{eqn:first_order_ECM_equations} with optimization variables $\bm{\xi} = (Z,V_1)$ and $\bm{\Theta} = (\alpha_{R_0,0-3},\alpha_{R_1,0-3},\alpha_{C_1,0-3})$. As expected from the $\delta_y$ analysis in Sec. \ref{subscn:signal_richness}, the results improve with respect to those in Fig. \ref{fig:R1_interp}.

\section{Multi-rate MHE}
\label{scn:multiscaleMHE}

This section proposes a further refinement of the \textit{standard} MHE described in Sec. \ref{scn:rich_adapt}, based on an analysis of model \eqref{eqn:first_order_ECM_equations} dynamics time scales. As reported in Sec. \ref{scn:model}, the modified HPPC cycle lasts for 60s, with a charge and discharge period of 10s and 20s. This cycle directly affects the $V_b$ dynamics, as it relates to the first-order system described by $R_1(Z)$ and $C_1(Z)$. However, as reported in Sec. \ref{scn:rich_adapt}, the SOC has slower dynamics in the order of hours. Indeed, this separation in the model dynamics frequencies has been neglected so far, considering a buffer $\text{Y}_k$ with $N_{T_s}=20$. However, the estimation performance would benefit from the faster dynamics too. Thus, we propose a modified version of the \textit{standard} MHE where the measurement buffer $\text{Y}_k$ is built with a non-uniform down-sampling factor $N_{T_s}$ within the time window. More specifically, the down-sampling selection within the time window is the following: 

\begin{itemize}
    \item Buffer length: $N = 30$ (\cite{Aeyels})
    \item Samples 1-5: $N_{T_s} = 1$. This sub-window captures the fast dynamics of the modified HPPC cycle
    \item Samples 6-30: $N_{T_s} = 20$. This sub-window captures the slow SOC dynamics
\end{itemize}

\begin{figure}
    \centering
    \includegraphics[width=9cm]{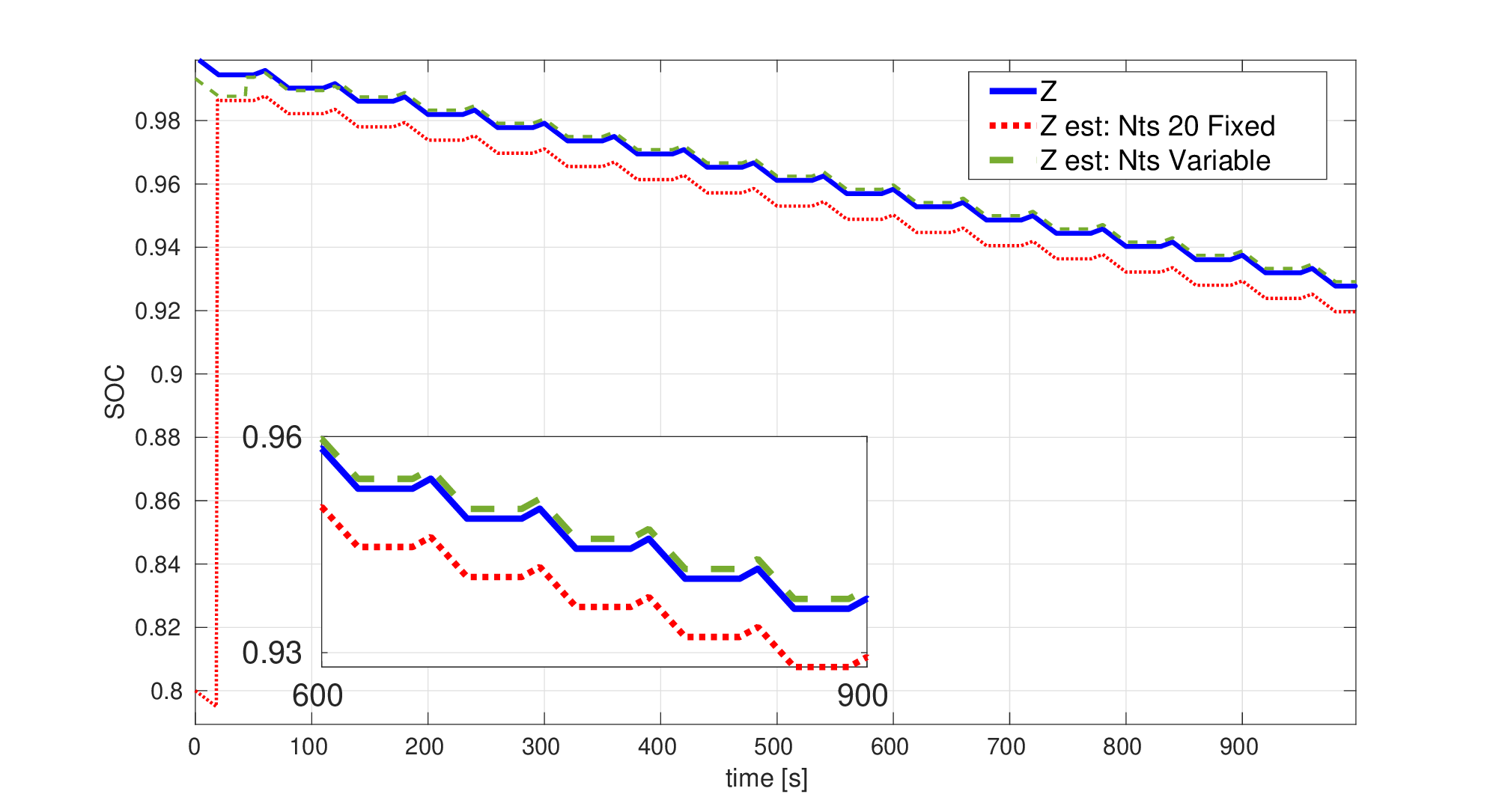}
    \caption{SOC estimation comparison between \textit{standard} MHE and \textit{multi-rate} MHE.}
    \label{fig:StandardMultiSOCcompare}
\end{figure}

Results in terms of SOC estimation are presented in Fig. \ref{fig:StandardMultiSOCcompare}, showing an improvement in the performance. Note that each method waits for the measurement buffer to be filled. That's why the estimation converges at different times in fig. \ref{fig:SOCVb_parallel}. As expected, the performance on the estimation of the parameters does not improve with respect to the \textit{standard} MHE with $N_{T_s}=20$. Note that the estimation error does not reach zero, as measurement noise affects $\text{Y}_k$; thus, the solution to \eqref{eqn:minproblem_barrier} is not exact. Indeed, the lower the noise, the better the estimation.

\section{Parallel MHE}
\label{scn:parallelMHE}

The main drawback of the \textit{standard} and \textit{multi-rate} MHE approaches described in Sec. \ref{scn:rich_adapt} and Sec. \ref{scn:multiscaleMHE} lies in the fact that the computational effort increases with $N_{T_s}$; this would happen as more model integrations are required on each optimization step. This is highlighted in Tab. \ref{tab:Nts_time_results} where the average optimization time is reported for the \textit{standard} MHE with $(N=30,N_{T_s}=1)$,$(N=30,N_{T_s}=20)$, and \textit{multi-rate} MHE $(N=30,N_{T_s}=\text{Variable})$. All these configurations have optimization variables $\bm{\xi}=[Z,V_1],\bm{\Theta} = \alpha_{(\cdot,0-3)}$. 

\begin{table}[h!]
	\centering
	\begin{tabular}{|c|c|c|}
		\hline
		Algorithm & Setup & Single iteration \\
		\hline
		Standard MHE & $N=30,N_{T_s}=1$ s  & 0.28 s \\
		Standard MHE & $N=30,N_{T_s}=20$  & 4.5 s \\
		Multi-rate MHE & $N=30,N_{T_s}=$Var  & 4.2 s \\
		
		\hline
	\end{tabular}
	\vspace{0.2em}
	\caption{\label{tab:Nts_time_results} \textit{Standard} and \textit{multi-rate} MHE computation times for different configurations.}
\end{table}

Indeed the computational time of both the \textit{standard} MHE with $N_{T_s} = 20$ and the \textit{multi-rate} MHE exceed the $N_{T_s} = 1$ configuration by an order of magnitude. This result wouldn't be an issue if the \textit{standard} MHE with $N_{T_s} = 20$ was used, as the time interval between two subsequent measurements is fixed to 20s. However, this would become an issue if \textit{multi-rate} MHE was considered, as the first five measurements in the buffer are sampled at a distance of 1s only.\\
Thus, this section proposes further refining the MHE approach, aiming to simultaneously provide a slow-paced model parameters estimation and a fast-paced SOC estimation. The general idea is presented in Fig. 9, and we refer to it as \textit{parallel} MHE. A slow $\text{MHE}_s$ is used to estimate the model parameters $\bm{\Theta} = (\alpha_{R_0,0-3},\alpha_{R_1,0-3},\alpha_{C_1,0-3})$, while a second fast $\text{MHE}_f$ provides only the estimation of $\bm{\xi} = (Z,V_1)$. Every time the $\text{MHE}_s$ updates the model coefficients $\bm{\Theta}$, the very same are also updated in the $\text{MHE}_f$ model. Another option, not furtherly investigated in this work, would consider differently structured observers instead of the fast-paced MHE, e.g., Kalman-based filters. However, note that the slow-paced MHE should be kept, as the parameter estimation is possible thanks to the long-term window of data considered.

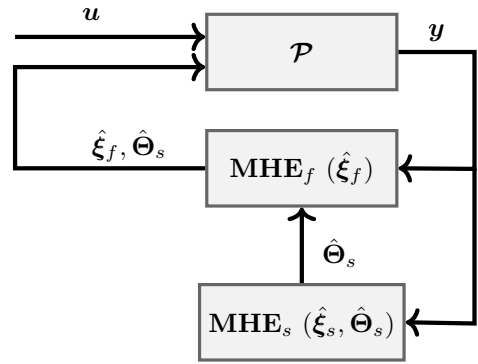
\begin{figure}[h!]
        \label{fig:par}
	\centering
	\begin{tikzpicture}[
			squarednode_black/.style={rectangle, draw=black!60, fill=black!5, very thick},
			circlenode/.style={circle, draw=black!60, fill=black!5, very thick},
			node distance=0.5cm]
			
		\node[squarednode_black,minimum width = 2.5cm,minimum height = 1cm] (P) {$\bm{\mathcal{P}}$};
		\node[squarednode_black,minimum width = 2.5cm,minimum height = 1cm] (Mfast) [below=of P] {$\text{\textbf{MHE}}_{f} \ (\hat{\bm{\xi}}_f)$};
		\node[squarednode_black,minimum width = 2.5cm,minimum height = 1cm] (Mslow) [yshift=-0.5cm,below=of Mfast] {$\text{\textbf{MHE}}_{s}\ (\hat{\bm{\xi}}_s,\hat{\bm{\Theta}}_s)$};

		\draw[black, ultra thick, ->] (P.east) -- ([xshift=1cm]P.east) node [xshift=-0.5cm,above] (TextNode) {$\bm{y}$} -- ([xshift=1cm,yshift=-1.55cm]P.east)  -- (Mfast.east);
		\draw[black, ultra thick, ->]  ([xshift=1cm,yshift=-1.55cm]P.east) -- ([xshift=1cm,yshift=-3.6cm]P.east) -- (Mslow.east);
		\draw[black, ultra thick, ->]  (Mslow.north) node [xshift=0.5cm,yshift=0.4cm] (TextNode) {$\hat{\bm{\Theta}}_{s}$} -- (Mfast.south);
		\draw[black, ultra thick, ->] (Mfast.west) node [xshift=-1cm,yshift=0.3cm] (TextNode) {$\hat{\bm{\xi}}_f,\hat{\bm{\Theta}}_{s}$} -- ([xshift=-2.5cm]Mfast.west) -- ([xshift=-2.5cm,yshift=-0.2cm]P.west) -- ([yshift=-0.2cm]P.west);
		\draw[black, ultra thick, ->] ([xshift=-2.5cm,yshift=0.2cm]P.west) node [xshift=1cm,yshift=0.3cm] (TextNode) {$\bm{u}$} -- ([yshift=0.2cm]P.west);

	\end{tikzpicture}	
	\caption{\centering Parallel MHE scheme}
\end{figure}

\medskip
The following setup has been considered: 

\begin{itemize}
    \item $\text{MHE}_s$: \textit{multi-rate} MHE with $N^s=30$ and $N_{T_s}^s$ as in Sec. \ref{scn:multiscaleMHE}. The optimized variables are $\xi_s = (Z_s,V_{1,s})$ and $\bm{\Theta}_s = (\alpha_{R_0,0-3},\alpha_{R_1,0-3},\alpha_{C_1,0-3})$.
    \item $\text{MHE}_f$: \textit{standard} MHE with $N^f=30,N_{T_s}^f=2$. The optimized variables are $\xi_f = (Z_f,V_{1,f})$. Moreover, $\bm{\Theta}_f$ is updated to $\bm{\Theta}_s$ every $N_{T_s}^s$s.
\end{itemize}

\begin{figure}[h!]
    \centering
    \includegraphics[width=8cm]{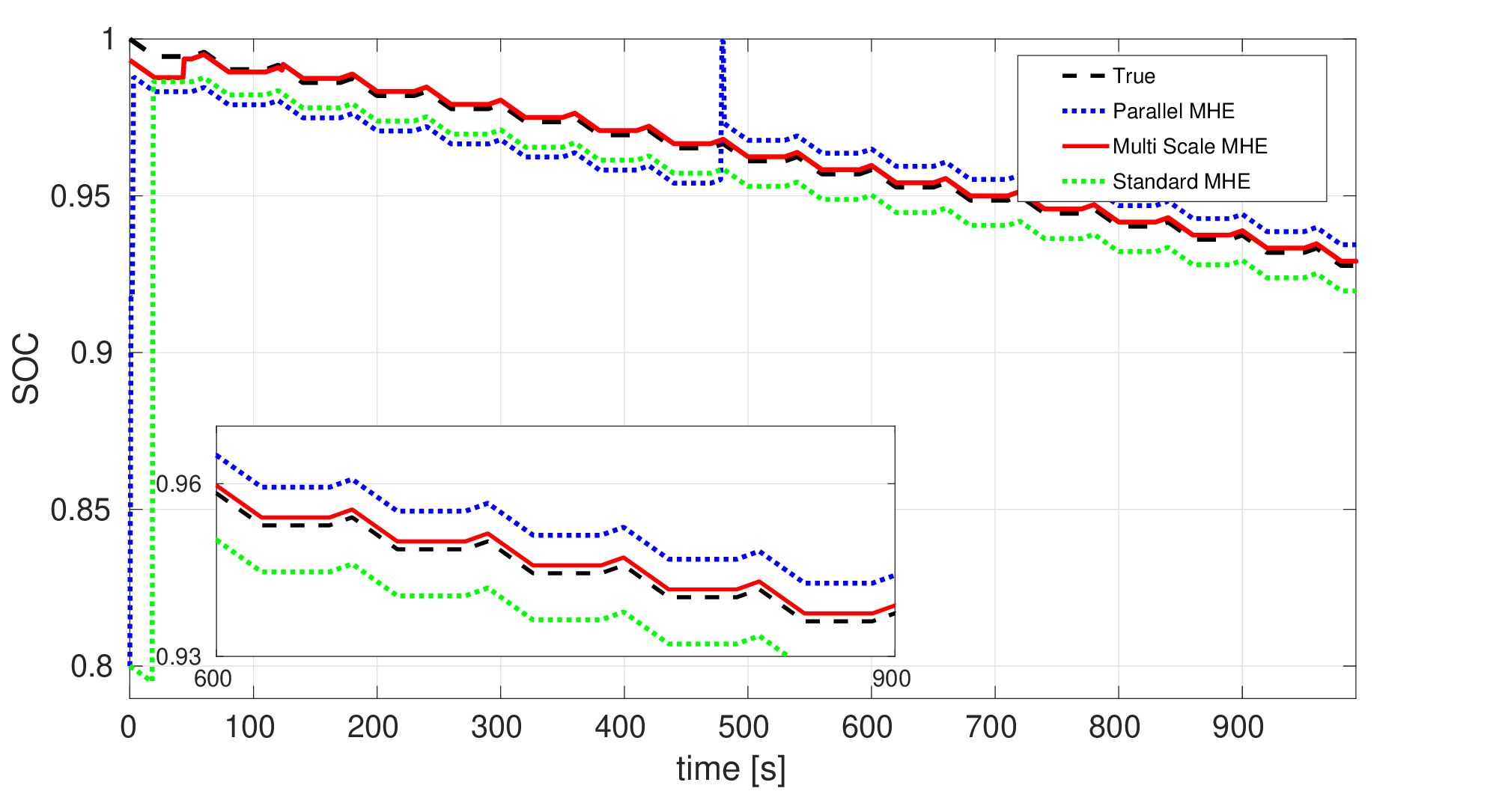}
    \caption{\centering SOC and $V_b$ estimation with \textit{parallel} MHE.}
    \label{fig:SOCVb_parallel}
\end{figure}

\medskip
The $\text{MHE}_f$ timescale is now comparable with the HPPC input cycle. Results in terms of SOC and $V_b$ estimation are presented in Fig. \ref{fig:SOCVb_parallel} where \textit{standard} MHE, \textit{multi-rate} MHE, and \textit{parallel} MHE are compared. Indeed, the \textit{parallel} MHE slightly worsens the performance with respect to the \textit{multi-rate} MHE but improves with respect to \textit{standard} MHE. As far as the parameters are concerned, the estimation follows the trend presented in Sec. \ref{scn:multiscaleMHE}. 

\medskip
Again, different algorithms imply different convergent times. Indeed, the \textit{parallel} MHE has to wait until the \textit{multi-rate} buffer is filled, which happens at time $t = 480$ s, hence the spike in the plot of fig. \ref{fig:SOCVb_parallel}. Indeed, the most significant improvement lies in the computational time, which drops significantly for the $\text{MHE}_f$, as described in Tab. \ref{tab:MultiMHE_time_results}. Thus, the \textit{parallel} MHE can be implemented online, justifying its usage against the \textit{multi-rate} MHE. 

\begin{table}[h!]
	\centering
	\begin{tabular}{|c|c|c|}
		\hline
		Algorithm & setup & Single iteration \\
		\hline
		$\text{MHE}_f$ & $N=30,N_{T_s}=2,(Z,V_1)$ s  & 0.1 s \\
		$\text{MHE}_s$ & $N=30,N_{T_s}=\text{Var},(\alpha_{\cdot,0-3})$ s  & 4.2 s \\
		\hline
	\end{tabular}
	\vspace{0.2em}
	\caption{\label{tab:MultiMHE_time_results} \textit{Multiple} MHE computation times.}
\end{table}

To conclude, the \textit{parallel} MHE allows an online implementation of the observer, reaching a good estimation accuracy both in terms of SOC and model parameters.
\section{Conclusion}
\label{scn:conclusion}

This paper considers a Moving Horizon Estimator (MHE) to estimate the SOC and the model parameters. We highlight that estimation performance is strongly affected by the window buffer length and the sampling time. This paper exploits the \textit{signal variability} index introduced in \cite{Oliva01} to provide a procedure to select the MHE window size and (down-)sampling time ($N_{T_s}$), maximizing the estimation performance. Moreover, we move even further by introducing a \textit{multi-rate} MHE algorithm capable of catching dynamics with different time scales, resulting in significant performance improvements. 
Lastly, a further refinement is introduced, namely a \textit{parallel} MHE capable of estimating at the same time slow and fast plant dynamics. This solution keeps a good estimation precision, making the algorithm implementable in real-time. Numerical results are provided all over the paper to support the validity of the analyses and the proposed solutions.\\
Indeed, this study uses a simplified ECM model and does not consider process noise, which will be investigated in future works. Furthermore, future developments will consider the \textit{filtered} MHE proposed in \cite{Oliva01} to increase the estimation performance even more by exploiting signal filtering. We  also plan to model parameter's dependence on temperature and the battery's State of Health (SOH). 


\vspace{-5pt}
\bibliography{bibliography.bib}

\end{document}